\input{aipcheck}

\documentclass[
    ,final            
  ]
  {aipproc}

\layoutstyle{6x9}
\newcommand{\ave}[1]{\left\langle #1 \right\rangle}

\newcommand{\order}[1]{ \mathcal{O} \left( #1 \right) }

\begin{document}

\title{Scaling of elliptic flow in heavy ion collisions}


\author{Giorgio Torrieri$^a$,Barbara Betz$^b$, Miklos Gyulassy$^c$}{
  address={$\phantom{A}^a$FIAS and $\phantom{A}^b$ITP, JW Goethe Universitat, Frankfurt,Germany, and $\phantom{A}^c$Department of physics, Columbia University, New York}
}

\begin{abstract}

The common interpretation of $v_2$ in heavy ion collisions is that it is produced by hydrodynamic flow at low transverse momentum  and by parton energy loss at high transverse momentum.   In this talk we discuss this interpretation in view of the dependence of $v_2$ with energy, rapidity and system size, and show that it might not be trivial to reconcile these models with the relatively simple scaling found in experiment
\end{abstract}

\maketitle
\section{Introduction}
Two experimental results of heavy-ion collisions have been subject to many theoretical and phenomenological investigations \cite{sqgp}:  One is the observation of a significant suppression of high- $p_T$ particles, ``jet quenching'', the other one is the observation of an azimuthal dependence of the particle spectra on the reaction plane $\phi_{0n}$ at both high and low momenta, the ``elliptic flow''. The elliptic flow,
$v_2$, is parametrized as the second Fourier component of the transverse momentum distribution of the produced particles 
\begin{equation}
\label{v2def}
 \frac{dN}{dp_T dy d\phi} =  \frac{dN}{d p_T dy}\left[1+ \sum_{n=1}^\infty 2 v_n(p_T) \cos(\phi-\phi_{0n}) \right]\,.
\end{equation}
The interpretation of the first finding is generally thought to be that the matter produced in heavy-ion collisions is ``opaque'', with a large energy loss per unit length of fast particles \cite{lpm,Baier:1996sk}.
The second finding has been interpreted in terms of the ``perfect fluid'', the hypothesis that matter in heavy-ion collisions has an extremely low viscosity \cite{fluid1,olli}.
Hence, initial anisotropies in configuration-space density of the collision area will be transformed into anisotropies in the collective flow of matter.

It is important to emphasize that $v_2$ is present at all values of momentum, but has different origins at different momenta if the consensus outlined above is correct.
Elliptic flow at low-$p_T$ is due to the hydrodynamic evolution of the system while elliptic flow at high-$p_T$ is thought to be due to opacity, since partons emitted in the reaction plane loose less energy than partons emitted perpendicular to the reaction plane due to the shorter distance traveled.
Both are thought to be a dynamical response of the primary asymmetry present in heavy-ion collisions.
While the scale delimiting these two regimes is assumed to be the average $p_T$ of the system, $\ave{p_T} \sim \order{0.5-1}$ GeV (with the tomographic regime actually appearing at $\order{3-4}\times \ave{p_T}$), the way these two mechanisms combine at realistic $p_T$ is not entirely clear.
In addition, it might be possible that {\em some of} both elliptic flow and jet suppression are not generated due to medium but due to initial state effects.  Strong color dynamics at the parton saturation scale (the ``Color Glass Condensate'') has been shown to exhibit some jet suppression and elliptic flow \cite{lappi,cgc,nara}.

Phenomenologically distinguishing between different models, even at a given energy, is not so trivial because {\em every} model has quite a few undetermined fit parameters.    Hence, for instance, it is not as yet clear whether jet energy loss proceeds by weakly coupled or strongly coupled \cite{ches1,ches2} jet-medium dynamics, and we are far from understanding at what energy, if ever, do these effects significantly change.

One important experimental finding which can be used to clarify these questions is the discovery of a {\em scaling} in elliptic flow across different energies, system sizes and centralities, when the data is plotted against transverse momentum $p_T$ rapidity $y$,pseudo-rapidity $\eta$ and transverse multiplicity density $(1/S)(dN/dy)$, and eccentricity $\epsilon$.   Some experimental observations which help us define this scaling are \cite{qmposter}:
\begin{itemize}
\item The dependence of $v_2/\epsilon$ at mid-rapidity on only the transverse multiplicity density $(1/S)(dN/dy)$ across all available energies, system sizes and centralities \cite{cms}
\item The ``limiting fragmentation'' of $v_2$ in rapidity \cite{whitephobos}
\item The approximate independence of $v_2(p_T)$, in a given centrality class, on energy \cite{scanv2paper}, system size \cite{lacey} and rapidity \cite{brahmsscaling}
\end{itemize}
Note that the transverse area $S$ and the multiplicity $\epsilon$ are {\em theoretical} parameters, necessitating either a Glauber or a Color-Glass event-by-event Montecarlo simulation.   Further in this work we shall see how to cast some of this scaling in a purely experimental form.

One way to parametrize all this experimental data, at all energies and rapidities, is 
\begin{equation}
v_2(p_T) = \epsilon(b,A)F(p_T) 
\phantom{AA},\phantom{AA} \ave{v2} =  \epsilon(b,A) \int dp_T F(p_T) g\left( p_T,\ave{p_T}_{y,A,b,\sqrt{s}} \right) 
\end{equation}
Here, $\epsilon$ is the eccentricity (dependent, in a Glauber parametrization, somewhat weakly on energy, strongly on system size and centrality), $g(p_T)$ is 
the distribution in transverse momentum, approximately characterized by one parameter (the average momentum $\ave{p_T}$ or equivalently, the slope $T$), which in turn seem to depend, across rapidity $y$, center of mass energy $\sqrt{s}$ and centrality on just the initial density, in the Bjorken formula $\frac{1}{S}\frac{dN}{dy}$.
$F(p_T)$ seems to be a universal function, independent of both energy and eccentricity.

These are purely experimental statements, with no theoretical overlay, restating the results \cite{cms,whitephobos,scanv2paper,lacey,brahmsscaling} in mathematical form.  As such, they are ``as good as the error bar'', and a thorough scan in energy, system size and rapidity might in the coming years discover violations (some violation of the $p_T$ dependence can be seen at low $p_T$ in \cite{scanv2paper}).
Taking all this as an established fact, however, is an extraordinarily strong constraint, since ``typically'', for a complicated dynamical model (as non-linear hydrodynamics overlayed with jets inevitably is), generally does not factorize in {\em any} way, $v_2$ is simply $v_2(x_i)$ (where $x_i=\left\{ \sqrt{s},b,A,y,p_T \right\}$) and any element of $\partial^2 v_2 /\partial \ln x_i \partial \ln x_j$ is non-negligible

The first scaling, but {\em not} the universality of $F(p_T)$, was predicted in \cite{heiselberg} under certain assumptions (no phase transition) within a {\em weakly coupled} model (the Knudsen number $\sim 1$, so one interaction per degree of freedom per lifetime).   Either of the scalings have not been thoroughly explored in either hydrodynamics or tomography.   In this talk, we shall give some qualitative limits to the applicability of each scaling.
\section{\label{hydro}Some comments on scaling in hydrodynamics}
It has long been pointed out, both by heuristic arguments \cite{mescaling1,mescaling2} and explicit simulations \cite{heinzscaling} that the patterns above pose a problem for the hydrodynamic interpretation of $v_2$.
Close to the hydrodynamic limit, one expects that $v_2$ is
\begin{itemize}
\item Approximately $\propto \epsilon$ since $v_2 (\epsilon=0 )=0$ and $\epsilon$ small and dimensionless
\item Approximately $\propto c_s(T)$ since $v_2 (c_s=0 )=0$, and the dimensionless $c_s$ tracks the equation of state
\item $v_2$ is maximum for ideal hydro.   Since the Knudsen number $Kn$, quantifying the ratio of the mean free path to the system size, is small and dimensionless, $v_2 \sim v_2^{ideal} (1-Kn)$.    In turn, the Knudsen number is related to the viscosity over entropy density $\eta/s$ as well as the system size $R$, $Kn \sim \eta/(sTR)$
\item Finally, $v_2^{ideal}$ is a highly non-linear function of the lifetime $\tau_{life}$, $v_2^{ideal} \sim v_2(\tau_{life}/\tau_0 \rightarrow \infty) \times f(\tau_{life}/\tau_0)$, which can be numerically shown to be monotonically saturating, $\sim f(\ave{p_T}) \tanh(...)$ in a Cooper-Frye \cite{cf,heinzcf} freezeout. $\tau_{life}$ is in turn related to the freezeout temperature and energy density $T_f,e_f$.
\item For $p/\pi \ll 1$ and isothermal freeze-out, $\tau_{life}/\tau_0 \sim (e_{0}/e_f)^{4 \alpha}$, with $\left. \frac{1}{3} \right|_{bjorken} < \alpha < \left. 1 \right|_{hubble}$ depending on how ``three dimensional'' is the flow.  This relation becomes more complicated, but qualitatively similar, for systems at high chemical potential.
\end{itemize}
In summary, elliptic flow in the hydrodynamic limit should scale as
\begin{equation}
\frac{v_2}{\epsilon} \sim c_s  f \left( \frac{1}{T_f^3 \tau_0 R^2}\frac{dN}{dy}  \right) \left( 1 - \order{1} \frac{\eta}{s}  \frac{1}{ T R} \right)  
\end{equation}
It is clear that only $\order{Kn}$ terms mix intensive quantities such as the energy density $e$ with extensive ones such as the size $R$.    $\order{Kn^0}$ ``ideal'' terms, except for the initial time $\tau_0$, depend purely on intensive quantities, giving rise to scaling between systems of different sizes.   As we will see, this is {\em not} true for high $p_T$.
For low $p_T$, no scaling violation is seen in experimental data, giving a bound for Knudsen number compatible with $\eta/s=0$ \cite{nantes1}, albeit with large error bars (which still need to be computed), at all energies.
Moreover, the lack of scaling of $\tau_0$ is troubling since, by causality, it can locally {\em only} depend on energy and the local intensive parameters.
Just by dimensional analysis, it is difficult to see how   It can be constant w.r.t. energy, since $\tau_0 \sqrt{s} \sim \order{1}$.  Landau hydrodynamics would imply $\tau_0 \sim 1/\sqrt{s}$, while a CGC type initial condition would most likely give a logarithmic dependence since $\tau_0 \sim Q_s^{-1}$.   Either, however, would lead to unobserved systematic scaling violations.   The only way to avoid these is to assume $\tau_0$ is of the order of the local mean free path at equilibrium, and hence gets reabsorbed as a function of the entropy density, $\left( \sim (1/S)(dN/dy) \right)^{-1/3}$ for an ideal Equation of state.

Additionally, the Cooper-Frye formula \cite{cf,heinzcf} leads to a {\em non-universal} $F(p_T)$.    To show this, it is sufficient to expand it azimuthally in eccentricity \cite{cf,heinzcf}
 \begin{equation}
v_2(p_T) \simeq \int d\phi \cos^2 (2\phi) \left[  e^{ - \frac{\gamma\left(E-p_T  u_T \right)}{T}} \left(1 - p_T \Delta \frac{dt}{dr}  +  \frac{\gamma \delta u_T (\phi)  p_T }{T}  + \order{\epsilon^2,Kn} \right) \right]
\end{equation}
As long as $\delta v_T,  \Delta \frac{dt}{dr}  \sim \epsilon s^0$, $v_2(p_T)$ is independent of $\sqrt{s}$.   This will be true in the limit where the hydrodynamic phase is ``long'', $\tau_f/\tau_0 \gg 1$, but will {\em not} be the case \cite{heinzscaling,olli} if the duration of the hydrodynamic phase $ \leq \epsilon R$, as is the case at lower energies.   The introduction of an ``iso-knudsen freezeout'' rather than an isothermal one, a physically reasonable scenario explored in \cite{duslingteaney}, should further break this scaling.
The reasons for this behavior go all the way to the qualitative description of how $v_2$ behaves in hydrodynamics: $v_2(p_T)$ and $p_T$ integrated $\ave{v_2}$ scale differently, because in hydro Fourier components of the transverse flow
$v_T(r,\phi)$ depend on lifetime $\tau_{life}$ differently:
\begin{description}
\item[$v_2(p_T)$] depends only on the 2nd Fourier component of $v_T(r,\phi) \sim \tanh(\tau_{life}/\tau_0)$
\item[$\ave{v_2}$] depends on both the 0th ($\ave{p_T}\sim  \left( \tau_{life}/\tau_0 \right)^\omega$) and 2nd component.   
\end{description}
Given these, making $v_2(p_T)$ independent of energy but varying $\ave{p_T}$ strongly at all energies in unnatural in hydrodynamic models.  Detailed simulations including chemical potential, however, are needed to determine when does $\tau_{life}$ become ``short'', and more experimental data might yield a breakdown of $v_2(p_T)$ scaling at lower energies.

A further consideration is in order regarding the breakdown in scaling in {\em particle species} \cite{scanv2paper}.
$v_2^i(p_T)$ ($i=\pi,K,p,\Lambda,...$) does {\em not} scale the same way by particle species as it does for all particles:   different particle species $v_2^i(p_T)$ are different at different energies, but the differences cancel out when total $v_2 \simeq \sum_i v_2^i(p_T) n_i(p_T)/(\sum_i n_i(p_T))$ is considered($n_i$ is the particle species abundance).
In a Cooper-Frye freezeout \cite{cf} there is no reason for such a cancellation between flow and hadrochemistry to happen.   Coalescence models, while they will also break coalescence {\em scaling} at lower energies \cite{dunlop,megreco}, also do not predict such behavior.   

We note, as a speculative suggestion, the fact that structure functions $f(x,Q^2)$ and fragmentation functions $D_{q \rightarrow i}(z,Q^2) $ naturally follow the scaling suggested by both the overlap of $v_2(p_T)$ and its breakdown by particle species, since both of these depend weakly on momentum exchange $Q^2$ but strongly on the rescaled variables $x=p_z^{parton}/E,z=p^{hadron}/p^{parton}$ \cite{qcdcoll}.  In structure functions, $x$ is absorbed into the longitudinal component of momentum, with $Q^2 \sim p_T^2$.  In fragmentation functions, $z \sim p_T$ but unitarity protects the effect of fragmentations on all hadrons, $\sum_i\int z dz D_{q \rightarrow i}(z,Q^2)=1$. Together, these lead to a particle species dependent $v_2^i(p_T)$ with $\sqrt{s}$, but a much weaker dependence of total $v_2(p_T)$.   The suggestion that perhaps $v_2(p_T)$ is not a reflection of flow at all, but of non-perturbative QCD response of geometry needs much more theoretical development, and has to contend with the difficulty of pQCD appearing at $\sqrt{s}=7.7$ GeV (a comparatively low energy, although much above the ``minijet'' 1 GeV scale). If the overlap of high $p_T$ with low $p_T$ $v_2$, examined in the next section, persists, scenarios like this might however need further development.
\section{Higher $p_T$ $v_2$: Scaling in the tomographic regime}
While the scaling we discovered can be, to a certain extent, understood in the hydrodynamic regime, the tomographic regime is widely expected to break it.  This is because in hydrodynamics flow, and hence $p_T$-correlations, is generated by density {\em gradients}.   In tomography ,it is generated by path length variations.   Hence, the role of ``size'', $\ave{R}$, which typically depends on system size as $\sim A^{1/3}$ and is weakly dependent on energy, is very different in the tomographic regime w.r.t. the hydro regime. As we saw in the previous section, ``extensive'' factors $\ave{R T}$ (size$\times$ temperature) in the hydrodynamic regime are suppressed by $\order{Kn}$, and hence vanish in the ideal hydrodynamic limit.   In the tomographic regime, $v_2 \sim \epsilon f(\ave{R T} )$ and the $\ave{R T}$ dependence is not suppressed by any small parameter.   The exact form of the function depends on the details of the model, but, as we shall see, it generally does not drop out when $\ave{R}$ and $\ave{T}$ are varied separately by scanning in centrality, system size and $\sqrt{s}$.     This is true in ``standard'' jet energy loss calculations, both weakly \cite{lpm,Baier:1996sk} and strongly coupled \cite{ches1,ches2}, where opacity is a smooth function of the entropy density.    It is even more true in models, such as \cite{liao}, where it strongly depends on the distance from the deconfinement transition.   Changes in the opacity parameter (parametrized by $\kappa$ in Eq. \ref{abceq}) suggested in \cite{betz2}, as well as changes in the structure function moments in $\sqrt{s}$, are also expected to make the scaling {\em worse}.

Of course, the transition between ``hydro'' and tomography is a smooth superseding rather than a ``turning on/off''.   Indeed, hydrodynamics and tomography can be defined in terms of the Knudsen number in momentum space:
 assuming the scattering cross section $\sigma$ depends on the exchanged momentum $Q$ as $\sigma \sim 1/Q^2$, and assuming momentum is much higher than temperature, the Knudsen number becomes
$\frac{l_{mfp}}{R} \sim \frac{p_T^2}{s R}$
Hence, the tomographic regime starts dominating when $Kn \sim 1$. For different energies and transverse densities the critical $p_T$ can be easily shown to be
\begin{equation}
 \frac{p_{T1}^{tomo}\left( \sqrt{s_1},b_1,A_1 \right)}{p_{T2}^{tomo}\left( \sqrt{s_2},b_2,A_2 \right)} \sim \left( \frac{S_2}{S_1} \frac{dN_1/dy}{dN_2/dy} \right)^{\omega}
\end{equation}
where $\omega=1/2$ for collisional-dominated equilibration but could increase to $3/2$ \cite{dusling} if radiative processes become important in the hydrodynamic regime.

The advantage of characterizing as ``hard'' hadrons with  $p_T \geq p_T^{tomo}$ is that this definition is independent of details such as the dynamics of production and fragmentation of fast hadrons (we do not have to call them ``jets'', which is problematic at low energies).   Experimentally, the fact that $p_T^{tomo}$ decreases with decreasing $N_{part},\sqrt{s}$ is advantageous.

$v_2$ in the tomographic regime of course continues to be a good observable.   In particular, it is independent of effects like Cronin effect and jet reconstruction, which become problematic at low energies:   At lower energies $R_{AA}$ diverges, due to kinematics, at {\em any} medium opacity.   Since kinematic effects, {\em by themselves}, do not depend between the hadron-reaction plane angle, however, $v_2$ does need opacity to be generated.    This makes comparing $v_2 (p_T>p_T^{tomo})$ at {\em different energies and system sizes}, and looking for scaling violations, an optimal probe of changes in opacity with temperature.
 In the next section we use the ABC model \cite{betz1,betz2} to put some limits on this scaling in a class of models where $v_2(p_T \gg \ave{p_T})$ is generated by tomography alone.  We shall make the case that an experimental investigation of this scaling is a useful test for energy loss models. 

\subsection{A theoretical investigation using the ABC Model}
Since the reaction plane was determined at low $p_T$, one can now eliminate the theoretical input $\epsilon,S$ from our observables by concentrating on $v_2(p_T)/\ave{v_2}$.   The denominator is the momentum-integrated $v_2$, which should include all eccentricity information.    If $v_2$ depends purely on gradients, of course, this ratio should be independent of both energies and system sizes.  The size dependence, however, should lead to a break in the scaling.

The ABC model is a simple parametrization which describes the energy loss of a ``fast'' particle ($p_T/T \ll 1$) in traveling ``large'' medium ($1/(T \tau) \ll 1$, $\tau$ is the propagation time).   If the parton is light and on-shell, energy loss models should give
\begin{equation}
-\frac{dE}{d\tau} = f(T,p_T,\tau) \simeq \kappa p^{a} T^{b} \tau^c + \order{\frac{T}{p_T},\frac{1}{T\tau}}
\label{abceq}
\end{equation} 
$a,b,c$ are a nice  phenomenological way of keeping track of \underline{every} jet energy loss model in certain limits.   
In a collisional dominated parton cascade $a=1,c=0$ \cite{bamps}, in a radiative dilute plasma (``Bethe-Heitler regime'') $a=1/3, c=0$, in a dense plasma (LPM regime) $a=1,c=1$) while in a ``falling string'' AdS/CFT scenario \cite{ches1,ches2} $a=1/3,c>2$.   

\begin{figure}
  \includegraphics[height=.2\textheight]{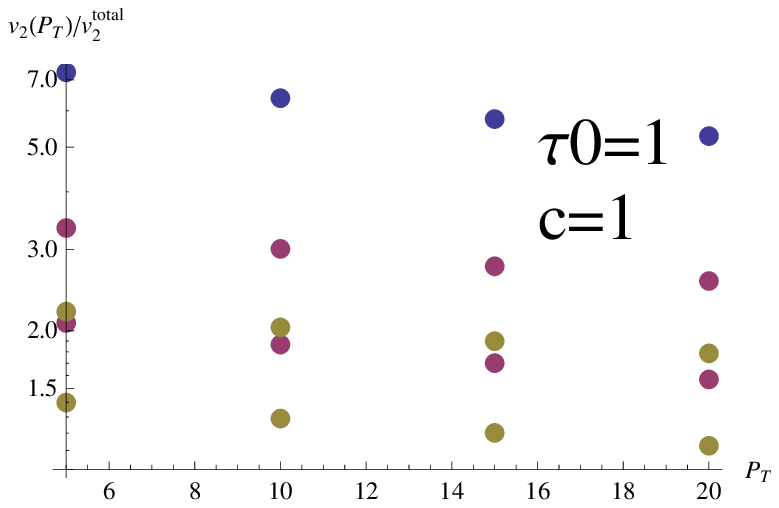}
  \includegraphics[height=.2\textheight]{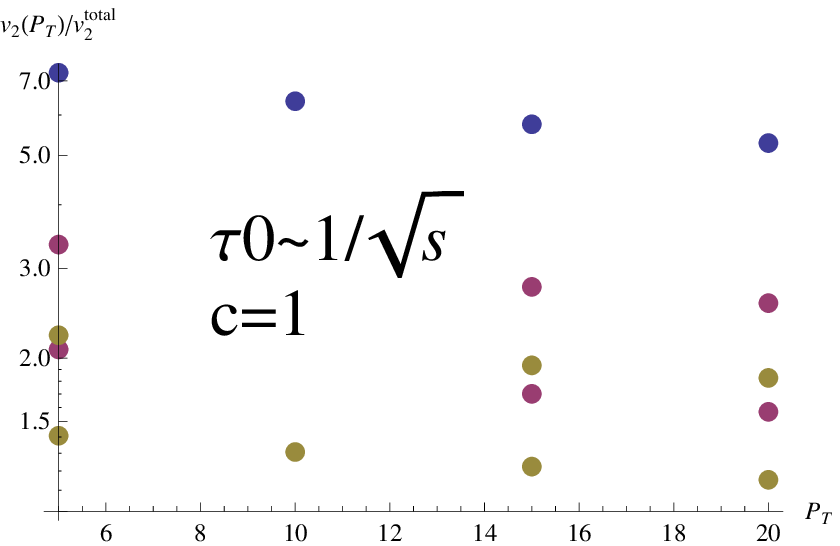}
  \includegraphics[height=.2\textheight]{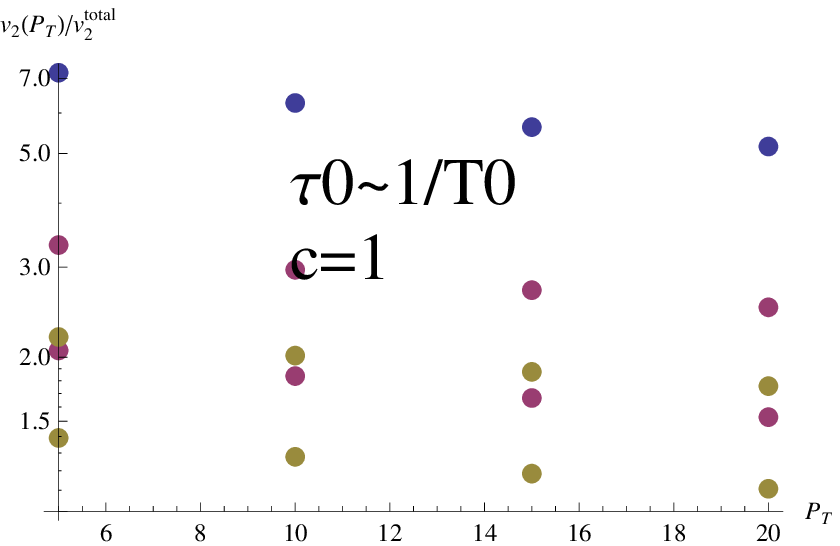}
\caption{Scaling plots assuming $z=1$, indicative of a pQCD dense plasma (LPM limit) energy loss, for RHIC Cu+Cu,RHIC Au+Au and LHC Pb+Pb \label{figlpm}}
\end{figure}

\begin{figure}
 \includegraphics[height=.2\textheight]{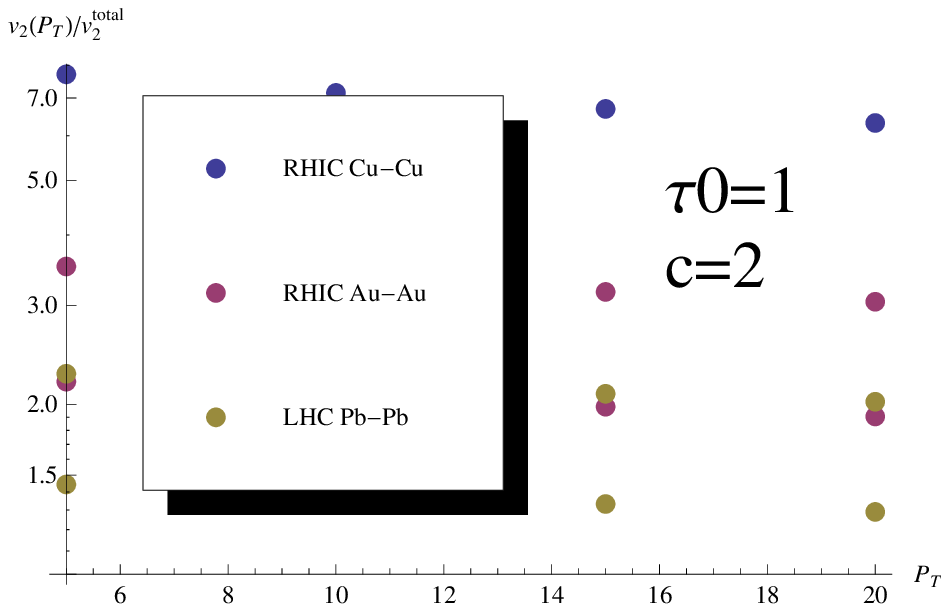}
  \includegraphics[height=.2\textheight]{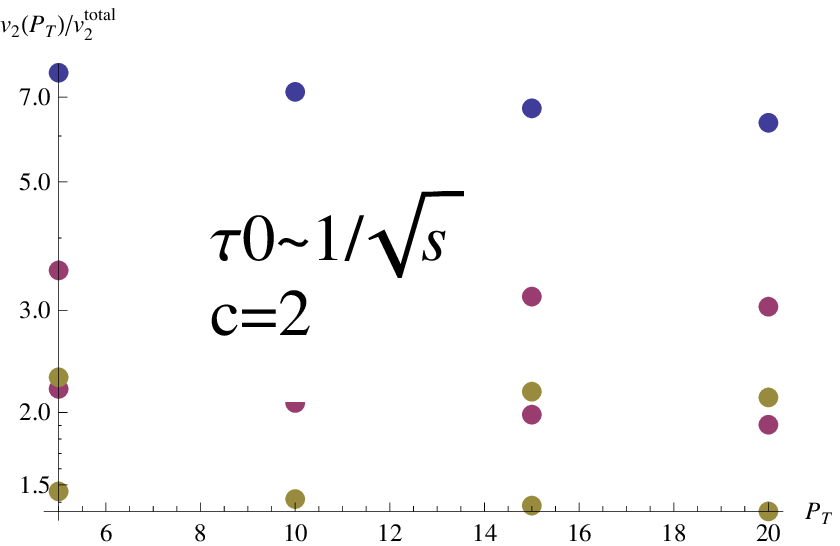}
  \includegraphics[height=.2\textheight]{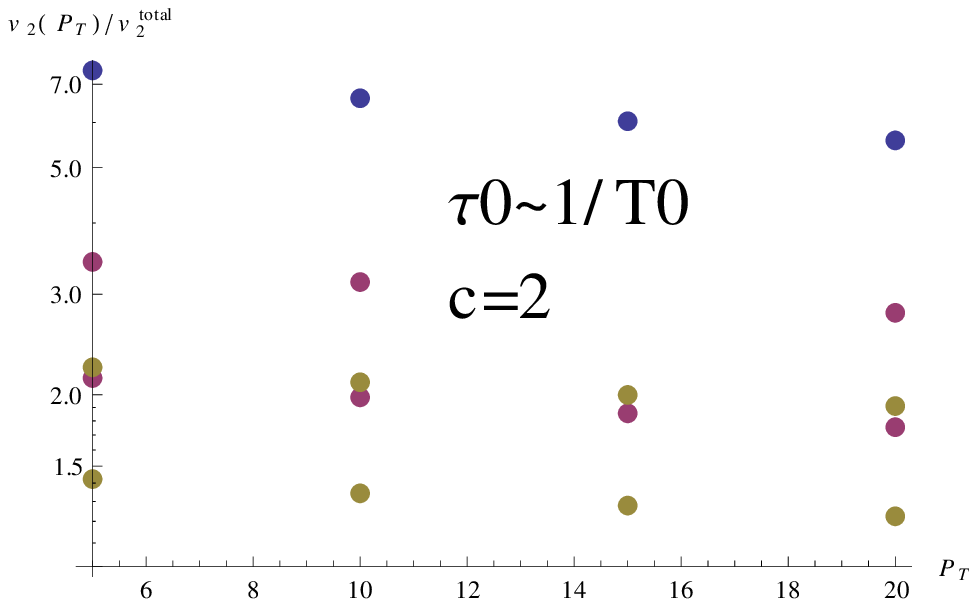}
\caption{Same as the previous figure, but $z=2$, indicative of AdS/CFT energy loss \label{figads}}
\end{figure}

\begin{figure}
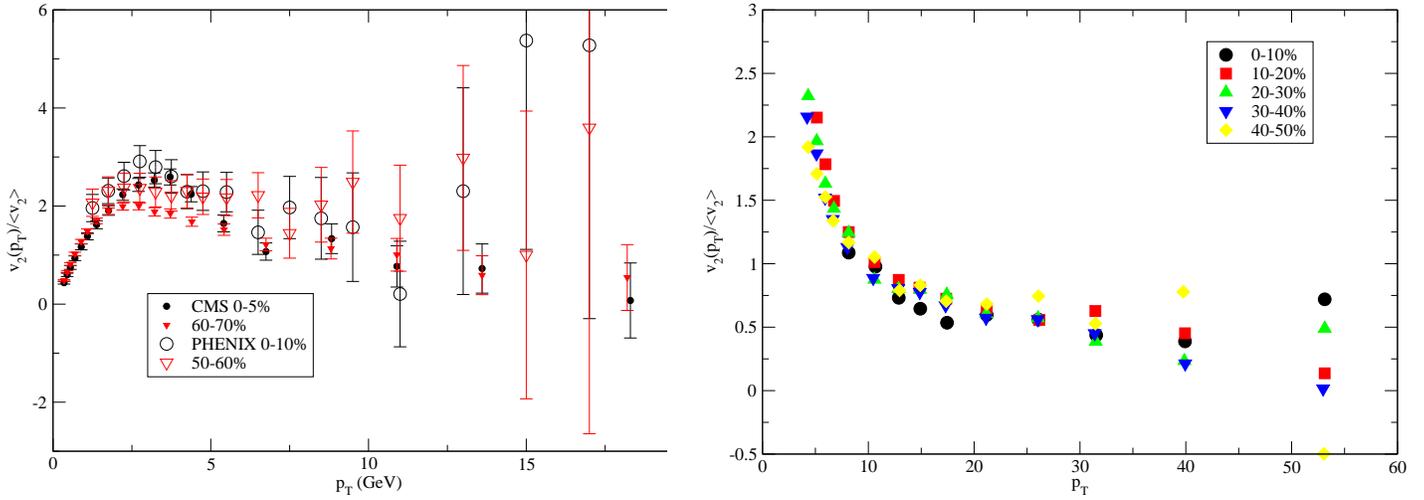

  \includegraphics[height=.3\textheight]{v2pt_cipanp_low.eps}
  \includegraphics[height=.3\textheight]{v2pt_cipanp_high.eps}
  \caption{\label{cmsexp} Experimental $v_2(p_T)/\ave{v_2}$ for LHC and RHIC energies.  The left panel shows data from \cite{cms,lacey} at comparatively low $p_T$, the right panel from \cite{cmshighpt}.  The data on error bars for the data in the right panel is not yet made public, so we did not include it}
\end{figure}

We can now expand in ``empirical'' parameters $\left( \frac{\Delta E}{E}\right)^{\pm 1},\epsilon,$ and get
\begin{equation}
 v_2\left( p_T^{high} \right)  \sim \epsilon^{\alpha_{\epsilon}}  L^{\alpha_R(a,b,c) } p_T^{\alpha_{p_T}(a,b,c) } \left( \frac{dN}{dy} \right)^{\alpha_{dn/dy}}.
\end{equation}
with non-trivial exponents $\alpha_i$, calculated in each model in \cite{coming}.
The size parameter $L \sim \min\left( \ave{R},\tau_f-\tau_0\right)$, sampling the lifetime for some systems and the transverse size $\ave{R}$ for others.
We perform a numerical investigation with a background given by a longitudinally expanding ellipsoid, fitted to global multiplicity and system size
\begin{equation}
T(r_T,\phi,\tau,\tau_0) = T_0 \left(\frac{\tau_0}{\tau}  \right)^{1/3} \Theta\left( r - R\left( 1+ \epsilon \cos\left(2 \phi \right) \right) \right) 
\end{equation}
Where $R,\epsilon$ are scanned scanned across radii of Cu,Au,Pb, $\tau_0$ are chosen according to the assumptions described in section \ref{hydro} $\tau_0= 1fm,\sqrt{s}^{-1}, T_{0}^{-1}$, and $T_0$ is adjusted to reproduce multiplicity and all energies.    We then obtain $v_2(p_T)/\ave{v_2}$, the latter given by the experimental parametrization $\ave{v_2}/\epsilon = 4(10^{-3}) (1/S)(dN/dy)$.
The results, shown in Fig. \ref{figlpm} and \ref{figads} show that scaling fails, by a similar magnitude, in all models.
Currently, experimental data does not allow to make definite conclusions but, as shown in Fig. \ref{cmsexp} \cite{cms,cmshighpt,lacey}, modulo rather big error bars, the only scaling violation is seen at intermediate regions.    Scaling holds across different centralities up to well above $p_T \simeq 20$ GeV, and seems to break up only at $p_T \simeq 40$ GeV at the LHC.  While a systematic shift of the center is seen comparing RHIC and LHC, the error bars are way too big to attach any meaning to this conclusion.  This shift is {\em not } seen up to $p_T \simeq 3.5$, where, as  noted in \cite{lacey}, this scaling holds across RHIC Cu+Cu and Au+Au.

If such a result survives when the error bars decrease, and persists across Au+Au vs Cu-Cu at RHIC and Pb-Pb vs Ar-Ar at the LHC, it would be a very profound statement given Figs \ref{figlpm} and \ref{figads}.   It would make it inevitable that {\em non-tomographic effects} play leading roles in high $p_T$ $v_2$.
While such effects have been suggested , for both the initial stage \cite{nara} and final fragmentation \cite{colorflow}, it is generally not expected that they play a large role at high $p_T$.    A continuation of scaling might force us to revise such expectations, since causality makes it inevitable that {\em short-lived} processes scale as gradients (such as $\epsilon$) rather than extensive quantities such as $\ave{R^\alpha T^\beta}$
We therefore eagerly expect further scaling studies at high $p_T$ $v_2$ to test our expectation that $v_2 (p_T \gg \ave{p_T})$ is generated tomographically, and to use the scaling violation as a constraint on jet energy loss models.


\begin{theacknowledgments}
G.T. acknowledges the financial
support received from the Helmholtz International Center for FAIR within the framework of the LOEWE program (Landesoffensive zur Entwicklung
Wissenschaftlich-\"Okonomischer Exzellenz) launched by the State of Hesse.
We would like to thank Victoria Zhukova for helpful discussions and insights. 
\end{theacknowledgments}



\bibliographystyle{aipproc}   





\end{document}